# Incorporation of Velocity-dependent Restitution Coefficient and Particle Surface Friction into Kinetic Theory for Modeling Granular Flow Cooling

Yifei Duan and Zhi-Gang Feng*

*Department of Mechanical Engineering, UTSA, San Antonio, 78249, USA*

Kinetic theory (KT) has been successfully used to model rapid granular flows in which particle interactions are frictionless and near elastic. However, it fails when particle interactions become frictional and inelastic. For example, the KT is not able to accurately predict the free cooling process of a vibrated granular medium that consists of inelastic frictional particles under microgravity. The main reason that the classical KT fails to model these flows is due to its inability to account for the particle surface friction and its inelastic behavior, which are the two most important factors that need be considered in modeling collisional granular flows. In this study, we have modified the KT model that is able to incorporate these two factors. The inelasticity of a particle is considered by establishing a velocity-dependent expression for the restitution coefficient based on many experimental studies found in the literature, and the particle friction effect is included by using a tangential restitution coefficient that is related to the particle friction coefficient. Theoretical predictions of the free cooling process by the classical KT and the improved KT are compared with the experimental results from a study conducted on an airplane undergoing parabolic flights without the influence of gravity [Y. Grasselli, G. Bossis, and G. Goutallier, EPL (Europhysics Letters) **86**, 60007 (2009)]. Our results show that both the velocity-dependent restitution coefficient and the particle surface friction are important in predicting the free cooling process of granular flows; the modified KT model that integrates these two factors is able to improve the simulation results and led to a better agreement with the experimental results.

* Zhigang.Feng@utsa.edu

# I. INTRODUCTION

The study of granular flow is of interest in a wide variety of fields in fundamental and applied sciences, including industrial flows such as pneumatic conveying and fluidized bed reactors, and environmental flows such as sand dunes and snow avalanches. Granular matter under rapid flow conditions is most commonly modeled as a continuum phase. Kinetic Theory (KT) supplemented with numerical simulations is considered to be one of the best tools to describe the behavior of rapid granular flows [1-3]. Most of these KT models have been derived for dilute flows of smooth, frictionless particles [4-6], which are essentially extensions of the classical KT of non-uniform gases [7]. However, there is one important difference between granular particles and gas molecules: kinetic energy is conserved in gas molecule collisions but dissipated in particle collisions. The dissipation of energy in granular particle collisions is due to the particle inelasticity, which is measured by the coefficient of restitution $e$. Most of the KT models assume the coefficient of restitution $e$ for a specific granular material is a constant and independent of the particle impact velocities [8-11]. Due to the macroscopic size of particles, external fields such as gravity would have a significant effect on granular flows. This makes it very difficult to experimentally investigate the flow behavior of granular materials due exclusively to particle collisions. Instead, the Discrete Element Method (DEM) is often used to verify the theoretical solutions in the absence of gravity [12]. Good agreements between theoretical predictions of the KT models and the DEM simulations have been reported [11,13,14], which is understandable since the same constant coefficient of restitution $e$ as implemented in the KT models has also been used in the DEM simulations. However, these DEM or KT models that are based on a constant $e$ fail to predict some of the most basic features of the experimental results [15,16].

It has been widely reported that the coefficient of restitution $e$ strongly depends on the impact velocity of a particle in collision [17-20]. To accurately measure $e$ at small impact velocities could be very challenging. Early experiments were made in the presence of gravity with impact velocities typically larger than 10 cm/s. The measured $e$ shows a monotonic decrease as the impact velocity increases [21,22]. Based on the data from [21], Lun and Savage [23] were the first to incorporate the velocity-dependent $e$ into the KT. They adopted an exponential decay function for $e$ to roughly match the experimental results at an impact velocity ranging from 100 cm/s to 250 cm/s. Due to the limited data for low impact velocity, at that time it was believed that the particle deformation was essentially elastic and the energy dissipation was small at very low impact velocities [24,25]. As a result, the fitting function of $e$ in their study predicts $e=1$ at very low impact velocities.

The effect of particle surface friction is also important for the KT of granular flows. Lun and Savage [26] considered this effect in their KT model by using a constant tangential restitution coefficient $\beta$. When $\beta$ equals -1, the particles are frictionless and there is no change in the tangential component of the relative velocity. On the other hand, when $\beta$ equals 1, the tangential component of the relative velocity reverses completely and the particles are said to be perfectly rough. However, it has been shown that the tangential restitution coefficient $\beta$ is not an independent parameter; it is related to the particle inelasticity and surface friction [27]. To consider the particle surface friction, sliding and sticking mechanisms must be distinguished in the binary collision model, and a relationship between $\beta$, the friction coefficient $\mu$ and the normal coefficient of restitution $e$, has to be established. Furthermore, the rotational degree of freedom also needs to be taken into consideration in the KT model; both translational and rotational granular temperatures should be employed to characterize the random velocity fluctuations of granular particles [28,29]. The widely used modification of the KT model that considers particles of small friction coefficient was developed by Jenkins and Zhang [30]. They employed the same structure in their model as the classical KT model for frictionless particles, but replaced the coefficient of restitution $e$ with an effective coefficient of restitution that accounts for the additional loss of translational fluctuation energy due to friction. However, their modification does not work for particles with large friction coefficients such as those used in the experiment [31] which have a friction coefficient $\mu$ =0.6. A few more KT models have been developed since then. These models mainly use collision integrals to produce new constitutive relations for rough spheres. In these models both particle friction and rotation were considered for energy fluxes without limitation on the friction coefficient [32,33]. The influence of these collisional parameters on the simulation results of a gas-solid bubbling fluidized bed has been investigated which showed improved predictions when compared with the experimental results [34,35]. However, these models require the inputs of the initial and boundary conditions of rotational granular temperatures that are usually unknown for most granular flow systems, so they have been rarely used in the literature. It must be pointed out that the collision models mentioned above are for binary collisions and valid only for rapid granular flows. For dense granular flows where network interactions dominate, collisional stress model such as the one proposed by Zhang [36-38] has to be employed. It is also found that the translations and rotations could be correlated when particles are rough [39]. Simulations show $\beta$ could be tuned to produce a huge distortion from the Maxwellian distribution function in some cases [40].

Grasselli, Bossis and Goutallier [31] and Tatsumi, Murayama, Hayakawa and Sano [41] experimentally investigated the granular flow cooling processes in microgravity. Large discrepancies between the KT predictions and their experimental results have been reported [31]. They also found that the coefficient of restitution $e$ decreases as the impact velocity becomes small, which is contrary to the conventional belief that *e* is close to one at small impact velocities and continuously decreases as the impact velocity increases. Their efforts to address these discrepancies include the use of *e* as a function of the normalized fluctuation energy in the dissipation rate expression of the KT model and a constant roughness coefficient $\beta$ that ranges from -1 to 1 to account for the particle roughness. With these modifications, they were able to slightly improve their KT model results. Nevertheless, the predictions of their KT model still don't match the experimental results well; the discrepancy in granular temperature is as large as 200% at the initial stage of the cooling process. We think there are two main factors that contribute to the discrepancy: the use of a constant tangential restitution coefficient $\beta$ and an inappropriate use of the restitution coefficient profile in their KT model. To address these issues, a KT theory considering both tangential restitution coefficient $\beta$ and the friction coefficient $\mu$ should be adopted. Also, instead of employing a granular temperature dependent *e* in the KT model, an impact velocity-dependent *e* at particle level should be used and the expression of *e* should be incorporated into the derivation of the Boltzmann kinetic equation.

In this paper we investigate the large discrepancies between the theoretical prediction of the existing KT models and the experimental results and develop modified KT models for the free cooling process of the granular flows. One of the modifications is to incorporate the velocity-dependent coefficient of restitution *e* directly into the Boltzmann kinetic equation to derive the translational fluctuation energy dissipation rate. Unlike the velocity-dependent *e* profile proposed by Lun and Savage [23], the present *e* profile considers the adhesive forces at slow impact velocities as supported by the recent experimental studies and has a much smaller value at low impact velocities. We also examine two different approaches that incorporate the effect of particle surface friction into the KT models, which are named as Model I and Model II. Both models are derived based on the exact rates of translational and rotational fluctuation energy dissipation calculated by Herbst, Huthmann and Zippelius [42]. Model I is the extension of the KT model proposed by Jenkins and Zhang [30]. It determines the rotational granular temperature by assuming rotational energy dissipation rate is minimal, and the frictional effect could be absorbed into an effective restitution coefficient $e_{eff}$. Compared to the original model which is limited to small $\mu$ and does not consider $\beta_0$ as an input, the present Model I takes both $\beta_0$ and $\mu$ into consideration and it can be applied to a system with large $\mu$. On the other hand, Model II which is proposed by Herbst, Huthmann and Zippelius [42] considers the rotational energy dissipation and translational energy dissipation separately by solving their coupled equations. Results from these two models are compared to show that Model I is able to produce results that are comparable to Model II. Finally, we incorporate the velocity-dependent $e$ into Model I and Model II to study the free cooling process of a granular flow. The simulation results are compared with the results from the existing KT models as well as the experimental data [31].

## II. VELOCITY-DEPENDENT RESTITUTION COEFFICIENT

The coefficient of restitution $e$ is introduced to conveniently model particle collisions. Despite $e$ cited as a constant in many studies, early experiments have shown that $e$ could depend on the impact velocity for a given granular material [43,44]. This phenomenon was explained by the fact that the collision force depends on a combination of factors, including the elastic deformation at low impact velocities and the increased energy dissipation due to the plastic deformation at high impact velocities. The effect of the velocity dependent $e$ on the KT models has been analyzed by Lun and Savage [23], who adopted a varying $e$ that decays exponentially with the increasing impact velocity. However, contrary to the previous finding that $e$ increases at small impact velocity [15,18,20,45], recent experiments show that for spheres as large as a few millimeters, the restitution coefficient *e* sharply decreases when the impact velocity becomes small. This new finding was further explained by the existence of van der Waals attraction at relatively low surface energies for typical grain materials. Many granular systems have particle collisions with a small impact velocity typically below 20 cm/s, the van der Waals adhesion between the flattened parts of a particle's surface can lead to a reduced restitution coefficient at these low impact velocities [19,46].

In the experimental study of granular cooling process by Grasselli, Bossis and Goutallier [31], the authors attempted to consider the velocity-dependent *e* by incorporating it into the fluctuation energy dissipation rate of the KT model, implying that the use of *e* is based on the granular temperature or normalized fluctuation energy of granular materials. This cannot be accurate since *e* is defined for each individual particle collision and therefore, has to be a function of the impact velocity of two particles involved in the collision. In the present KT model, the velocity-dependent *e* is used to derive the dissipation term directly from the Boltzmann kinetic equation by integrating the velocities of all the particles in collisions.

The choosing of restitution coefficient *e* is very important in the KT modeling; it is the primary factor that determines the rate of dissipation of granular flows. The simplest experiment to measure *e* would be to drop a sphere onto a flat horizontal plate with the help of gravity, and then determine the velocities both before and after the collision. However, the impact velocities for this kind of experiments are typically larger than 1 m/s. Small velocities may not be easily achieved because of gravity. Consider dropping a bead at a height of 1 cm, its impact velocity when it strikes the plate could reach 44 cm/s in free fall without considering air drag. It will require a much smaller height in order to produce a small impact velocity. However, the accuracy of measurement deteriorates as the height decreases. Another method to study two-particle collision at small velocities is to suspend the two particles on a pendulum, each particle held by a string, then release one or both spheres from a certain height. This setup allows particles to collide at very low velocities when the strings are long enough. A more sophisticated approach, which only recently became possible, is to conduct the experiments in a microgravity environment. Without gravity, there will be no constraints on the motion of particles and extremely low impact velocities can be achieved. In Fig.1, the experimental results at low impact velocities from two different approaches, one using pendulum [20] and the other using microgravity [31], are compared. The diameters of steel beads used in the experiments are 3 mm and 2 mm, respectively. The data collected from these two experiments show a similar profile of velocity-dependent $e$. As the impact velocity changes from 25 cm/s to 100 cm/s, the restitution coefficient $e$ approaches 0.9, which is the value of the restitution coefficient of steel cited in many studies. However, both experiments show a rapid decrease of $e$ when the impact velocity drops below 25 cm/s. Considering that the two sets of data were obtained by two different methods yet they match each other very well, they provide a strong evidence to support the existence of adhesive forces when the surface energies are small and the impact velocity is low.

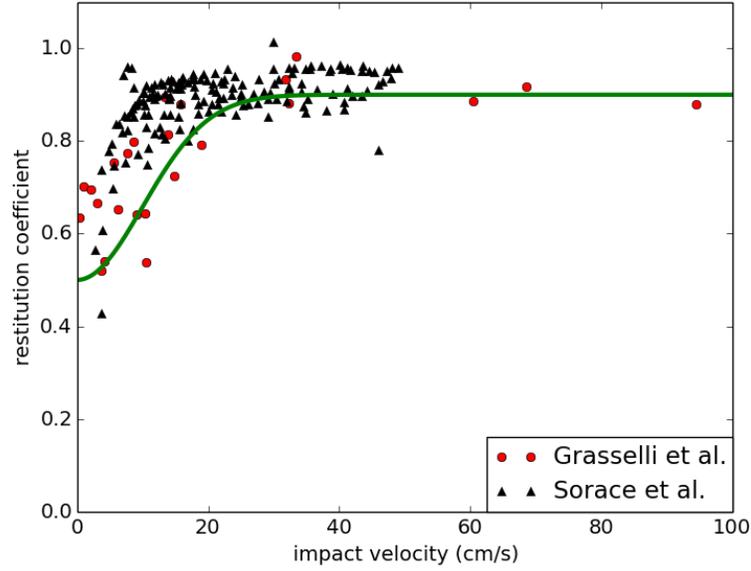

FIG. 1. Normal restitution coefficient *e* vs. the normal impact velocity. Soild circles show experimental results under microgravity condition from Grasselli, Bossis and Goutallier [31]; solid triangles show results using pendulum from Sorace, Louge, Crozier and Law [20]; solid line is the fitting result given in Eq. (1).

As explained by the Johnson, Kendall, and Roberts (JKR) theory [46], the adhesive forces between particles could cause *e* to decrease at low impact velocities. Some models that consider adhesive forces have successfully explained these experimental results [17,19,20]. However, those models require more specific material properties such as the van der Walls surface energy and viscous relaxation time that are hard to measure experimentally, and the expressions for *e* are usually much more complicated. Consider a collision between two identical spheres, each with a pre-collision velocity $c_1$ or $c_2$, and $c_{12} = c_1 - c_2$ is the relative velocity. The unit-vector connecting the center of two spheres is denoted by $k$. In the present study we chose a velocity-dependent *e* profile for steel particles by fitting the experimental data shown in Figure 1:

$$e = e_0 \left[ 1 - A \exp\left(-\left(c_{12} \cdot k / V_a\right)^2\right) \right] \tag{1}$$

Here, $e_0$ =0.9, $V_a^2$ =200cm²/s² and $A$ =0.4. This expression will be used in our KT model later on. It must be pointed out that the expression in Eq. (1) is only for systems of low granular temperature and low particle velocity since it is based on impact velocities below 100 cm/s. For systems of high granular temperature and high impact velocities, particle collisions may result in plastic deformation and reduce the coefficient of restitution, therefore different fitting profiles will be needed.

## III. KINETIC THEORY WITH LARGE SURFACE FRICTION

The original KT models are derived for frictionless and nearly elastic particles without particle rotational motion. However, granular materials are frictional and inelastic. The particles can rotate after a collision due to surface friction, so translational kinetic energy may be converted to rotational energy, affecting the dissipation rate of translational kinetic energy. Two kinds of collisions are defined in collision mechanics. The first one is sliding collision. In such a collision, the tangential collision force $F_{ij}^t$ exceeds the maximum friction force ($F_{ij}^t > \mu F_{ij}^n$), causing the particles to slide. Here, $F_{ij}^n$ is the normal component of the collision force, $\mu$ is the static friction coefficient. The tangential force arises from the Coulomb friction associated with the relative motion between the two spheres at the contacting surfaces. The other one is called sticking collision. In this case the tangential component of the collision force is below the maximum friction force ($F_{ij}^t < \mu F_{ij}^n$) and there is no relative motion between the contacting surfaces. In most KT models and hard sphere simulations, collisions are not resolved and the restitution coefficient $e$ is given as a constant. Similarly, earlier models involving particle rotations considered the tangential coefficient of restitution $\beta$ as merely a constant averaged over the entire range of sticking and sliding contracts [26,27]. The sliding and sticking mechanisms were later distinguished with the use of the friction coefficient $\mu$, the normal restitution coefficient $e$, and the tangential restitution coefficient $\beta_0$ for sticking collisions [47-49].

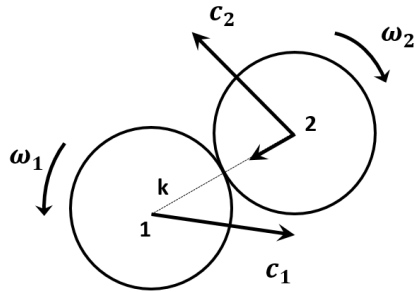

FIG. 2. A pair of colliding particles

Consider a collision between two identical spheres, each with a pre-collision velocity $c_1$ or $c_2$ and post-collision velocity $c'_1$ or $c'_2$. The relations between these velocities are

$$c'_1 = c_1 + \Delta c \tag{2}$$

$$c'_2 = c_2 - \Delta c \tag{3}$$

Similarly, the angular velocity $\omega$ changes after a collision follows

$$\omega'_1 = \omega_1 + \Delta\omega \tag{4}$$

$$\omega'_2 = \omega_2 + \Delta\omega \tag{5}$$

The relative velocity at the point of contact $g$ is given by

$$g = c_1 - c_2 + \frac{d}{2} k \times (\omega_1 + \omega_2) \tag{6}$$

According to the definition of restitution, we have

$$g' \cdot k = -e(g \cdot k) \tag{7}$$

$$g' \times k = -\beta(g \times k) \tag{8}$$

Based on Eqs. (2-8) and the conservation laws for linear and angular momentum, the translational and rotational velocity changes during a collision are equal to

$$\Delta c = -\frac{1}{2}(1+e)(c_{12} \cdot k)k - \eta k \times \left(c_{12} \times k + \frac{d}{2}\omega_{12}\right) \tag{9}$$

$$\Delta \omega = \frac{2\eta}{qd}\left[k \times c_{12} + dk \times (k \times \omega_{12})\right] \tag{10}$$

Where $c_{12} = c_1 - c_2$, $\omega_{12} = \omega_1 + \omega_2$, $\eta = [1+\beta]q/[2(1+q)]$, and $q = 4I/md^2$. Here $d$, $m$, and $I$ are the particle diameter, mass, and the moment of inertia, respectively. **k** is the unit vector directed from the center of particle 2 to particle 1. The tangential coefficient of restitution $\beta$ has to include both sliding collisions ($-1 \leq \beta_f \leq 0$) and sticking collisions ($0 \leq \beta_0 \leq 1$). The tangential restitution coefficient for sliding collisions $\beta_f$ depends on the impact angle at the point of contact $\theta$ between **g** and **k** as well as the surface friction coefficient $\mu$ [50], while the tangential restitution coefficient for sticking collisions $\beta_0$ should be constant. Whether a collision is of the sliding or sticking type depends on $\theta$. When this angle is greater than a critical angle $\theta_c$, a sliding collision occurs. On the other hand, a sticking collision take places if this angle is less than or equal to $\theta_c$. The effective tangential restitution coefficient is found to be,

$$\beta = \begin{cases} \beta_f & \theta \geq \theta_c \\ \beta_0 & \theta < \theta_c \end{cases} \quad (11)$$

And the critical angle $\theta_c$ can be obtained by forcing $\beta$ to be continuous

$$\tan \theta_c = \mu \frac{1+e}{1+\beta_0} \frac{1+q}{q} \quad (12)$$

where $\beta_f = -1 + \mu \frac{1+q}{q}(1+e)\cot\theta$, Eq. (12) has been proved to have a good agreement with the experimental results [49,51]. In general, if we consider $e$ as a function of impact velocity, both $\theta_c$ and $\beta_f$ could be affected. It is noted that at small impact velocity, there should be more sliding collisions (decreasing $\theta_c$) and less rotational energy dissipation (decreasing $\beta_f$) for sliding collisions compared to a system with constant $e$. However, when the results for constant $e$ and velocity-dependent $e$ of Eq. (1) are compared, as shown in Figure 3, we find the effect of velocity dependent $e$ on the particle rotational behaviors is insignificant. For example, if $\mu$ is small such as the case $\mu = 0.1$ in Figure 3, the change of $\theta_c$ at different impact velocities (when velocity-dependent $e$ is used) is small. By assuming $\theta_c$ as a constant, most of the collisions would be of sliding type since $\theta_c$ is small and the dissipation of rotational fluctuation energy mainly depends on $\beta_f$. However, the change of $\beta_f$ is also small between a constant $e$ and a velocity-dependent $e$ even at low impact velocities, as shown in Figure 3. On the other hand, if $\mu$ is large such as the case $\mu = 0.6$, the critical angle $\theta_c$ would also be large but the change of $\theta_c$ at different impact velocities is still insignificant when velocity-dependent $e$ is used. At large $\theta_c$, the portion of sliding collisions where $\theta > \theta_c$ would be small compared to sticking collisions. Even though $\beta_f$ can vary a lot during sliding collisions, the sticking collisions take $\beta_0$ as the tangential restitution coefficient which is a constant and $\beta_f$ would have little impact on the rotational energy dissipation rate. Therefore, as an approximation we will assume both $\theta_c$ and $\beta_f$ are independent of the impact velocity when we integrate the Boltzmann kinetic equations at the end of this section.

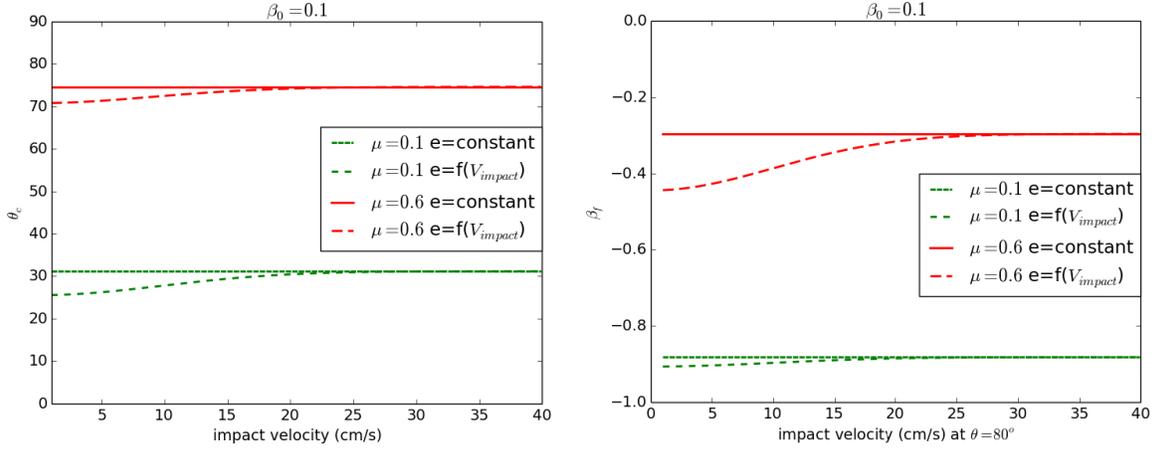

FIG. 3. The effect of velocity-dependent $e$ to the critical angle $\theta_c$ (left) and to the tangential restitution coefficient of sliding collisions $\beta_f$ (right).

To account for the energy dissipation due to particle rotations, we focus on the dissipation term of the Boltzmann kinetic equation. Since we only consider free cooling cases in this study, the collisional source term is given by

$$\chi(\Phi) = \frac{d^{N-1}}{2} \int_{\mathbf{c}_{12} \cdot \mathbf{k} < 0} \Delta\Phi (\mathbf{c}_{12} \cdot \mathbf{k}) f^{(2)}(\mathbf{c}_1, \boldsymbol{\omega}_1, \mathbf{r}_1, \mathbf{c}_2, \boldsymbol{\omega}_2, \mathbf{r}_2, t) d\mathbf{k} d\mathbf{c}_1 d\mathbf{c}_2 d\boldsymbol{\omega}_1 d\boldsymbol{\omega}_2 \qquad (13)$$

where $N$ is the number of dimensions, $\Delta\Phi$ represents the change of $\Phi$ during a collision, and $f^{(2)}$ is the coupled particle velocity distribution function,

$$f^{(2)}\left(\mathbf{c}_1,\boldsymbol{\omega}_1,\mathbf{r}-\frac{d}{2}\mathbf{k},\mathbf{c}_2,\boldsymbol{\omega}_2,\mathbf{r}+\frac{d}{2}\mathbf{k},t\right) = g_0 f_1 f_2 \left(1+\frac{d}{2}\mathbf{k}\cdot\nabla\ln\frac{f_2}{f_1}\right) \quad (14)$$

Here $\phi$ is the solid volume fraction; we choose $g_0(\phi) = \dfrac{2-\phi}{2(1-\phi)^3}$ for three-dimensional systems [52] and $g_0(\phi) = \dfrac{1-7\phi/16}{(1-\phi)^2}$ for two-dimensional systems [5], which is the expression for the radial distribution function at contact. In the homogeneous cooling state, by assuming the rotations and translations of particles are independent of each other, the unperturbed (zeroth-order) particle probability distribution function is of Maxwellian form for both the translational and rotational velocity fluctuations [53], that is

$$f(\mathbf{c},\boldsymbol{\omega},\mathbf{r},t) = n\left(\frac{1}{2\pi\Theta_T}\right)^{N/2}\left(\frac{I}{2\pi m\Theta_R}\right)^{N/2}\exp\left(-\frac{I\omega^2}{2m\Theta_R} - \frac{c^2}{2\Theta_T}\right) \quad (15)$$

The translational granular temperature is defined as $\Theta_T = \dfrac{\int c^2 f d\mathbf{c} d\boldsymbol{\omega}}{nN}$ and rotational granular temperature is defined as $\Theta_R = \dfrac{I\int \omega^2 f d\mathbf{c} d\boldsymbol{\omega}}{nmN}$, where $n$ is the particle number density.

The total change of translational energy in a collision equals $\Delta E = \dfrac{1}{2}m\left(c_1'^2 + c_2'^2 - c_1^2 - c_2^2\right)$. Based on Eq. (9), we find

$$\Delta E = \left\{-\frac{1}{4}m(1-e^2)(\mathbf{c}_{12}\cdot\mathbf{k})^2\right\} + \left\{m\eta(\eta-1)\left(|\mathbf{c}_{12}|^2 - (\mathbf{c}_{12}\cdot\mathbf{k})^2\right) + md^2\eta^2|\mathbf{k}\times\boldsymbol{\omega}_{12}|^2 + md\eta(2\eta-1)(\mathbf{k}\times\boldsymbol{\omega}_{12})\cdot\mathbf{c}_{12}\right\} \quad (16)$$

By letting $\Delta\Phi$ be $\Delta E$, Eq. (13) can be integrated to obtain the rate of dissipation of translational fluctuation energy per unit volume. The integration results could be divided into two parts, $\Gamma_T = \Gamma_{T1} + \Gamma_{T2}$, corresponding to the two terms in Eq. (16). The first term $\Gamma_{T1}$ is the energy dissipated due to the inelastic interactions in the normal direction; the second term $\Gamma_{T2}$ is caused by the interactions in the tangential direction. By assuming a constant normal restitution coefficient $e$, the exact rate of dissipation of the translational fluctuation energy calculated by Herbst, Huthmann and Zippelius [42] has the form

$$\Gamma_{T1} = -12\left[1-e^2\right]K(\Theta_T) \tag{17}$$

$$\Gamma_{T2} = -48\eta_0\left[\frac{1}{2}\left(\frac{\arctan\sigma}{\sigma}+\frac{1}{\sigma^2+1}\right)-\frac{1}{\sigma^2+1}(\lambda+1)\eta_0\right]K(\Theta_T) \tag{18}$$

Therefore, the 3-Dimensional (3D) translational fluctuation energy dissipation rate for homogeneous cooling is

$$\Gamma_T = -12\left[1-e^2\right]K(\Theta_T) - 48\eta_0\left[\frac{1}{2}\left(\frac{\arctan\sigma}{\sigma}+\frac{1}{\sigma^2+1}\right)-\frac{1}{\sigma^2+1}(\lambda+1)\eta_0\right]K(\Theta_T) \tag{19}$$

Similarly, the dissipation rate of the rotational fluctuation energy takes the form

$$\Gamma_R = -48\eta_0\left[\frac{1}{2}\left(\frac{\arctan\sigma}{\sigma}+\frac{1}{\sigma^2+1}\right)\lambda-\frac{1}{q}\left(\frac{1}{\sigma^2+1}\right)(\lambda+1)\eta_0\right]K(\Theta_T) \tag{20}$$

where

$$K(\Theta_T) = \frac{\rho\phi^2 g_0}{d\sqrt{\pi}}\Theta_T^{3/2} \tag{21}$$

Here, $\eta_0 = [1+\beta_0]q/[2(1+q)]$, $\sigma = (\lambda+1)^{1/2}\cot\theta_c$; $\phi$ is the solid volume fraction; $\lambda = \dfrac{\Theta_R}{q\Theta_T}$ is the ratio related to rotational granular temperature and translation granular temperature. For 2D systems, the translational and rotational fluctuation energy dissipation terms reduce to the followings [54]:

$$\Gamma_T = -4\left[1-e^2\right]K(\Theta_T) - 16\left[\frac{\eta_0}{2}\frac{1+\sigma^2}{(1+\sigma^2)^{3/2}} - \frac{\eta_0^2}{2}\frac{(1+\lambda)}{(1+\sigma^2)^{3/2}} - \eta_0^2\tan^2\theta_c\left(1-\frac{1+\frac{3}{2}\sigma^2}{(1+\sigma^2)^{3/2}}\right)\right]K(\Theta_T) \quad (22)$$

$$\Gamma_R = -16\left[\lambda\frac{\eta_0}{2}\frac{1+\sigma^2}{(1+\sigma^2)^{3/2}} - \frac{\eta_0^2}{2q}\frac{(1+\lambda)}{(1+\sigma^2)^{3/2}} - \frac{\eta_0^2\tan^2\theta_c}{q}\left(1-\frac{1+\frac{3}{2}\sigma^2}{(1+\sigma^2)^{3/2}}\right)\right]K(\Theta_T) \quad (23)$$

It must be pointed out that the above equations are derived for constant coefficient of restitution $e$ and they have to be modified for velocity-dependent $e$. Since $e$ is defined for an individual particle collision, it has to be a function of the impact velocity of two particles involved in the collision. This coefficient should be used to derive the rate of energy dissipation directly from the Boltzmann kinetic equation. According to Eq. (13), the translational energy dissipation rate is

$$\Gamma_T = \frac{d^{N-1}}{2}\int_{c_{12}\cdot k<0}\Delta E(c_{12}\cdot k)f^{(2)}(c_1,\omega_1,r_1,c_2,\omega_2,r_2,t)dkdc_1dc_2d\omega_1d\omega_2 \quad (24)$$

The velocity-dependent $e$ primarily affects the energy dissipated due to the inelastic interactions in the normal direction $\Gamma_{T1}$. The portion in the tangential direction $\Gamma_{T2}$ and rotational fluctuation energy dissipation rate $\Gamma_R$ remains the same since we assume a velocity-dependent $e$ would have insignificant effect on $\theta_c$ and $\beta_f$, as discussed earlier in this section. We only need to reintegrate the energy dissipation term $\Gamma_{T1}$ with a velocity-dependent $e$ given by Eq. (1),

$$\Gamma_{T1} = \frac{d^{N-1}}{2}\int_{c_{12}\cdot k<0}\left[\frac{1}{4}m(1-e^2)(c_{12}\cdot k)^2\right](c_{12}\cdot k)f^{(2)}(c_1,\omega_1,r_1,c_2,\omega_2,r_2,t)dkdc_1dc_2d\omega_1d\omega_2 \quad (25)$$

Details of the integration are provided in the Appendix. Overall the 3D translational fluctuation energy dissipation rate $\Gamma_T$ has the form

$$\Gamma_T = -12\left[1-e_0^2 + Ae_0^2 V_a^4\left(\frac{2}{\left(4\Theta_T+V_a^2\right)^2} - \frac{A}{\left(8\Theta_T+V_a^2\right)^2}\right)\right]K(\Theta_T) \tag{26}$$

$$-48\left[\frac{\eta_0}{2}\left(\frac{\arctan\sigma}{\sigma}+\frac{1}{\sigma^2+1}\right)-\frac{1}{\sigma^2+1}(\lambda+1)\eta_0^2\right]K(\Theta_T)$$

Similarly, the 2D translational fluctuation energy dissipation rate follows

$$\Gamma_T = -4\left[1-e_0^2 + Ae_0^2 V_a^4\left(\frac{2}{\left(4\Theta_T+V_a^2\right)^2} - \frac{A}{\left(8\Theta_T+V_a^2\right)^2}\right)\right]K(\Theta_T) \tag{27}$$

$$-16\left[\frac{\eta_0}{2}\frac{1+\sigma^2}{\left(1+\sigma^2\right)^{3/2}} - \frac{\eta_0^2}{2}\frac{(1+\lambda)}{\left(1+\sigma^2\right)^{3/2}} - \eta_0^2\tan^2\theta_c\left(1-\frac{1+\frac{3}{2}\sigma^2}{\left(1+\sigma^2\right)^{3/2}}\right)\right]K(\Theta_T)$$

To consider the rotational effect in the KT, we examined two models based on two different approaches: Model I is an extension of a simplified model originally proposed by Jenkins and Zhang [30] which includes the rotational effect into the dissipation rate of translational fluctuation energy; Model II is a model developed by Herbst, Huthmann and Zippelius [42] which solves the coupled Ordinary Differential Equations (ODEs) of Eqs. (19) and (20) and considers the rotational and translational fluctuation energy dissipation separately. Since both Model I and Model II consider the effect of particle surface friction, we would compare the new energy dissipation rates predicted by different models at constant *e*. The combination of both velocity-dependent *e* and particle surface friction will be studied and compared with experimental results in Section IV.

**A. Model I: Inclusion of the frictional effect by modifying translational granular temperature**

In Model I, the rotational granular temperature described in Eqs. (20) and (23) is not solved directly but the increased translational fluctuation energy dissipation caused by particle rotations is incorporated into the model by modifying the translational granular temperature. Similar to the assumption made in [30], we assume that the rotational fluctuation energy dissipation rate is very small (i.e., $\Gamma_R \approx 0$). This allows us to find $\lambda$ from Eqs. (20) and (23), which is treated as a constant for given particle properties. The modified translational fluctuation energy dissipation rates could then be obtained from Eqs. (19) and (22). Previous studies have shown that the extra translational fluctuation energy dissipation caused by the particle surface friction can be absorbed into an effective normal restitution coefficient [55]. Following this approach, several different effective restitution coefficient models have been proposed by different groups, including $e_{eff} = e - \frac{\pi}{2}\mu + \frac{9}{2}\mu^2$ [30], $e_{eff} = e - \mu + 2\mu^2(1+e)$ [56], and $e_{eff} = e - \frac{3}{2}\mu \exp(-3\mu)$ [11].

In Model I, the frictional effect is also included into the effective restitution coefficient. Compared to other models, the current Model I has two main advantages: a) there is no limitation on the friction coefficient $\mu$; and b) the tangential restitution coefficient for sticking collisions $\beta_0$ will be included. Figure 4 shows the comparison of the effective restitution coefficients between the Model I and other models found in the literature for a system with constant $e$=0.9.

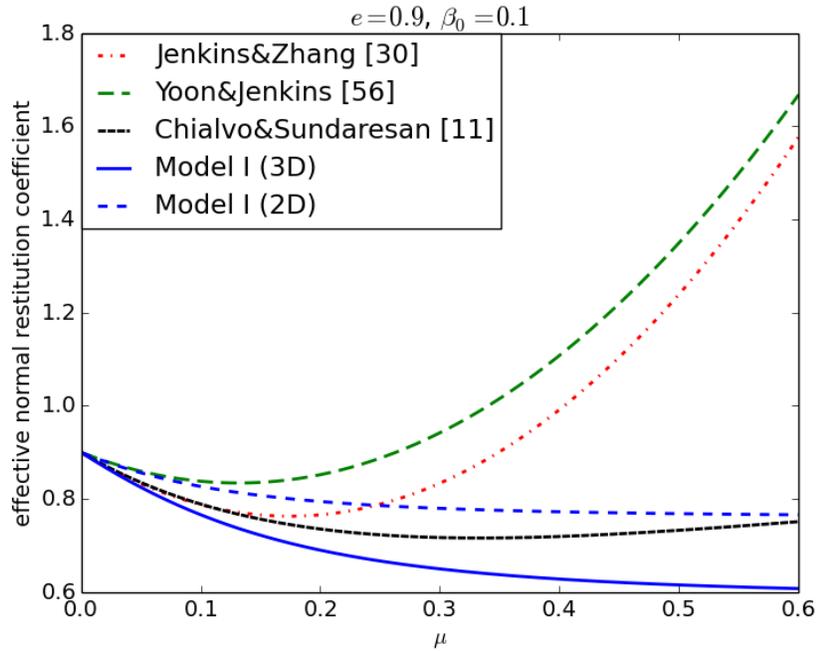

FIG. 4. Effective restitution coefficient derived from various models for a system with *e=0.9* and $\beta_0$ =0.1.

From Figure 4 we find that the original model by Jenkins and Zhang [30] and Yoon and Jenkins [56] only works at small restitution coefficient $\mu$; it starts to deteriorate as $\mu$ increases, eventually becomes unphysical when $\mu$ >0.4, resulting in $e_{eff}$ >1 and a prediction of the increase of granular temperature that is not possible. The fitting expression from the DEM simulation data by Chialvo and Sundaresan [11] matches well with the results of the present model for 3D systems when $\mu$ <0.05. However, their model shows a rising trend when $\mu$ >0.4, which is hard to explain since rougher particles should dissipate more energy. Unlike these original modifications that don't consider the tangential restitution $\beta_0$ for sticking collisions in the expression of $e_{eff}$, $\beta_0$ is incorporated into Model I. It is found that $\beta_0$ has a limited impact on the dissipation rate when $\mu$ is small. This explains why the original model [30] that ignores $\beta_0$ can still predict reasonable results [57]. However, as $\mu$ increases, the ratio of sticking collisions to sliding collisions also increases, resulting in a significant change of dissipation rate as seen in Figure 5, and the original model is no longer applicable.

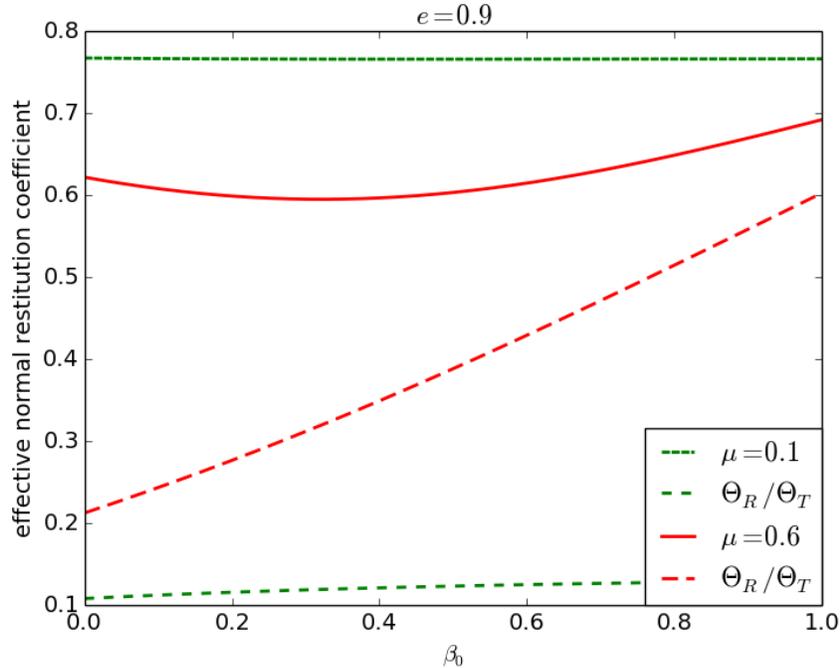

FIG. 5. The impact of tangential restitution coefficient $\beta_0$ on the effective restitution coefficient. The two dashed lines on bottom show the ratio of rotational granular temperature to translational granular temperature from Model I.

We also found the increased translational fluctuation energy dissipation rate caused by the frictional effect is less significant in a 2D system than in a 3D system. This finding is useful to help understand the discrepancy found between the theoretical predictions and the experimental measurements discussed in the next section. The experimental technique used in [31] was only able to measure two translational velocity components. In the experiment the particles were confined between two parallel plates with a distance of the particle diameter to make the system

2D. However, in order for the particles to move freely in the 2D plane, the distance between the two parallel plates would be slightly larger than the particle diameter, and there will be a small velocity component in the direction that is normal to the plane. Therefore, the experimentally measured translational granular temperature should be slightly less than the predictions of an ideal 2D system.

**B. Model II: Inclusion of the frictional effect by solving rotational granular temperature**

In this model we solve the two coupled ODEs given in Eqs. (22) and (23). Contrary to the Model I in which the ratio $\lambda$ is a constant and calculated by forcing $\Gamma_R = 0$, $\lambda$ in Model II is a variable determined by solving the additional governing equation for the rotational granular temperature. In addition, the initial rotational granular temperature $\Theta_{R0}$ is needed as an input in Model II. To understand the importance of different initial rotational granular temperature and compare it with Model I, we set up a free cooling granular case with a different initial rotational granular temperature $\Theta_{R0}$. Three cases are considered with the same initial translational temperature $\Theta_{T0}$ and different initial rotational temperatures of $\frac{\Theta_{R0}}{\Theta_{T0}} = 0$, $\frac{\Theta_{R0}}{\Theta_{T0}} = 0.25$ (with initial $\Gamma_R = 0$), and $\frac{\Theta_{R0}}{\Theta_{T0}} = 1$, respectively.

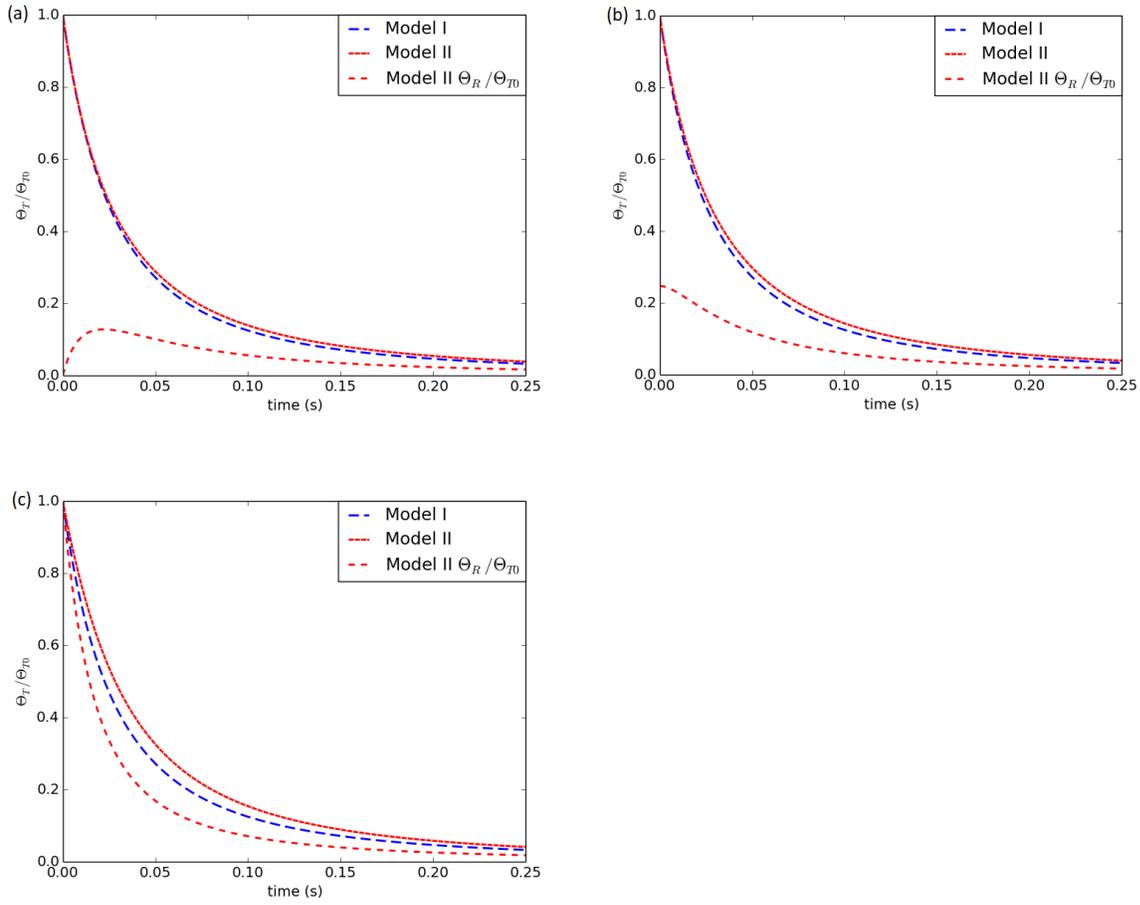

FIG. 6. The granular temperature vs. time for three cases with the same initial translational granular temperature but different initial rotational granular temperature: (a) $\frac{\Theta_{R0}}{\Theta_{T0}}=0$, (b) $\frac{\Theta_{R0}}{\Theta_{T0}}=0.25$ and (c) $\frac{\Theta_{R0}}{\Theta_{T0}}=1$. $\mu=0.6$ and $\beta_0=0.1$ are used in all the three cases. The two dashed lines on top show the translational granular temperature profiles predicted by the present models and the dashed line on bottom shows the dimensionless rotational granular temperature from Model II (Model I doesn't solve the rotational granular temperature, so the $\lambda$ which is related to the ratio of rotational to translational granular temperatures is kept at its initial value during the cooling process).

From Figure 6(a) we can see that the collisions are able to transfer more translational energy to rotational energy in the initial stage if the initial rotational granular temperature is low, causing an increase of the rotational granular temperature in a short period of time. After the rotational granular temperature reaches its maximum value, both rotational and translational granular temperatures start to decrease. The second case shown in Figure 6(b) predicts a translational granular temperature slightly higher than the first one due to the larger amount of initial rotational granular temperature. Finally, with $\frac{\Theta_{R0}}{\Theta_{T0}}=1$, Figure 6(c) shows that the large amount of rotational energy could be converted to the translational energy during collisions, making the translational granular temperature decay slowly. Overall, both Model I and Model II predict similar results, but their discrepancy becomes larger for systems with high initial rotational granular temperatures.

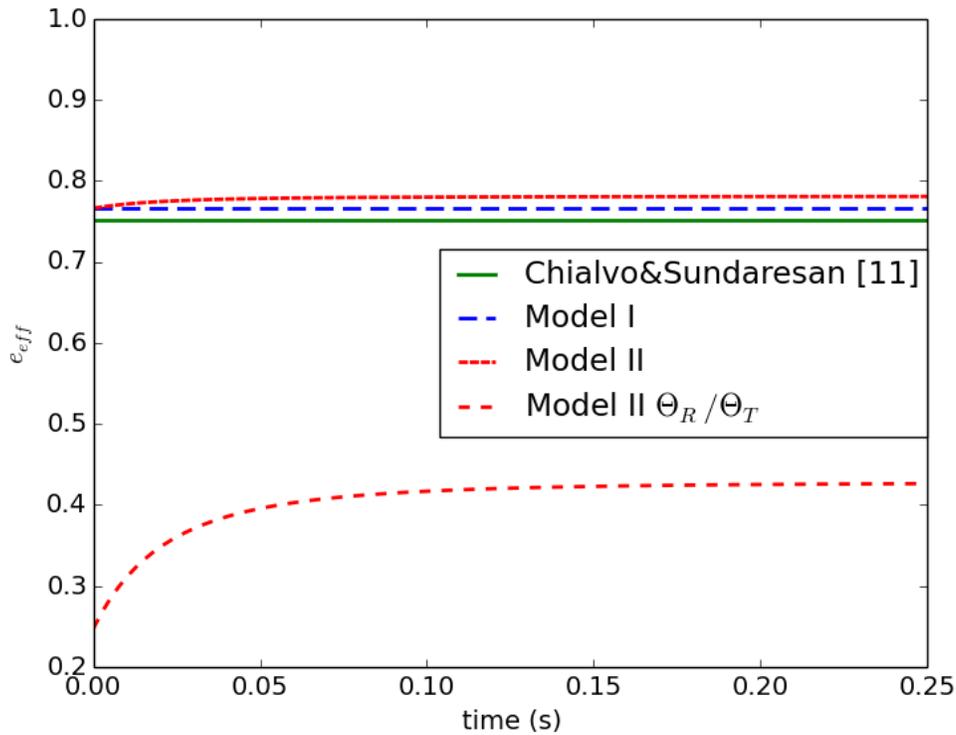

FIG. 7. The effective restitution coefficient vs. time during a cooling process. The dashed line on bottom shows the ratio of the rotational granular temperature to the translational granular temperature in model II during the cooling process for a 2D system with *e=0.9*, $\beta_0$ =0.1 and $\mu = 0.6$.

To better compare Model I and Model II, we make the initial rotational granular temperature $\Theta_{R0}$ the same for both cases. The case in Figure 6(b) was selected since its $\Theta_{R0}$ is calculated by forcing $\Gamma_R = 0$, which is the same as that in Model I. The resultant effective restitution coefficients are plotted in Figure 7. It is observed that when a system starts to cool down from its equilibrium state, $\lambda$ which is related to the ratio of the rotational to the translational granular temperatures increases and eventually reaches a constant value that is higher than the initial value. The similar trend was also reported in the previous work by Brilliantov, Pöschel, Kranz and Zippelius [39]. The increased $\lambda$ forces more rotational energy to be converted to translational energy. This explains the smaller translational fluctuation energy dissipation rate predicted by Model II when compared to Model I, since Model I simplifies the rotational effect by fixing the ratio $\lambda$ at its initial value, which is smaller. Overall both models predict an effective restitution coefficient that is very close to the value derived from the DEM simulation results by Chialvo and Sundaresan [11].

### IV. COMPARISONS WITH EXPERIMENTAL RESULTS

We test the two different KT models, Model I and Model II that consider the effect of particle surface friction from the last section, by applying them to the free cooling process of granular materials. The experiments were performed using a granular system composed of spherical particles under parabolic flights [31]. The particles were confined between two horizontally placed plates with a small gap slightly larger than the particle diameter. The plates were made of glass in order to cancel out the electrostatic effects and to minimize the friction between the particles and the walls. This setup only allows the free motion of particles that are parallel to the plane, resulting a 2D system. The granular system was initially vibrated and gradually came to rest due to the inelastic collisions. By analyzing the trajectories of the particles using a high-speed camera, their velocities could be determined, so the granular temperature of the systems could be calculated. Since the image analysis can only be performed in two dimensions, only two velocity components could be obtained. Such a setup used in the experiments with one layer of spherical particles could be treated as a 2D system filled with disks in KT models, and the area fraction instead of the volume fraction should be used in the calculations. For the steel beads, the restitution coefficient $e = 0.9$; the friction coefficient $\mu = 0.6$ and $\beta_0 = 0.1$, as provided in [31]. However $e$ may decrease at small impact velocities as shown in Figure 1 and a velocity-dependent $e$ of Eq. (1) will be used in our simulations

Figure 8 shows the translational granular temperature versus. time predicted by various KT models. The experimental results are also plotted for comparisons. It could be seen that the classical KT model which considers the system as 3D differs significantly from the experimental result; the discrepancy becomes even larger if we consider the system as a 2D system. Both Eq. (26) and (27), which treat the system as 3D and 2D respectively, under-predicts the energy dissipation rate and shows a much slower decay of the granular temperature. This is because the classic KT models does not account for the extra energy dissipation due to the particle surface friction and the increased inelasticity at a small impact velocity. If we consider the rotational motions with the given particle roughness while keeping the restitution coefficient $e$ constant, the three models including the model by Chialvo and Sundaresan [11], Model I, and Model II produce almost the same results. The initial rotational granular temperature in Model II is calculated by assuming the system at the equilibrium state. As shown in Figure 8, even though all three models have significantly improved their predictions in comparison to the classical KT models, they still significantly underestimate the decay rate of the translational granular temperature compared to the experimental results.

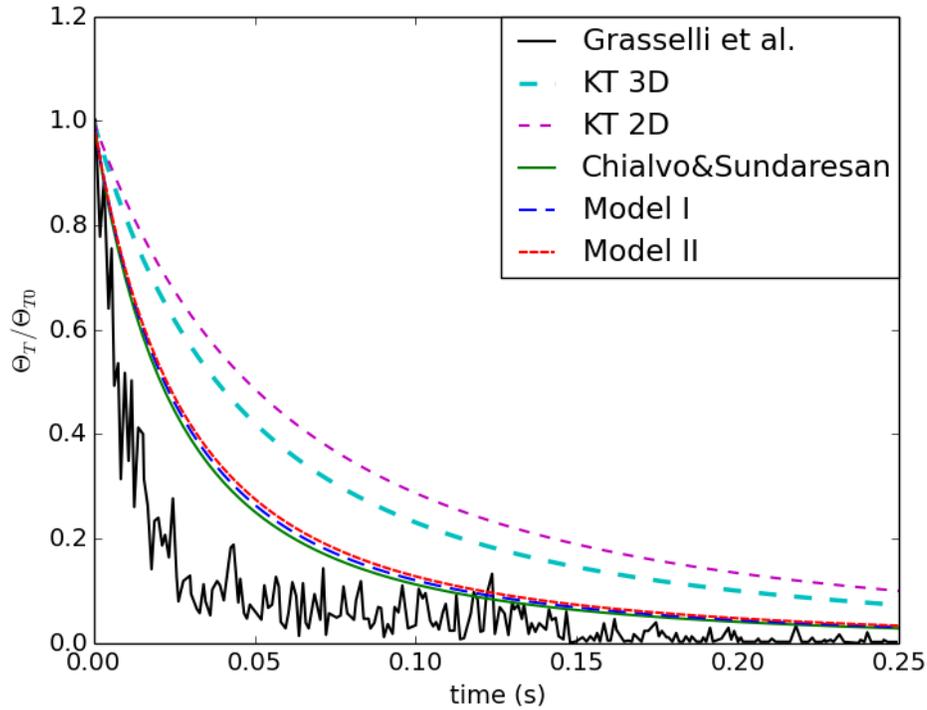

FIG. 8. Granular temperature along with the time during the free cooling process. The black line represents the experimental measurements from Grasselli, Bossis and Goutallier [31]. The two dashed lines on top show the theoretical predictions with the constant $e$ by the classical KT 2D and 3D models, the other lines show the results of the KT models considering surface friction by Chialvo and Sundaresan [11], the present model I and model II.

Figure 1 clearly shows that the restitution coefficient $e$ could be much smaller than the constant value $e = 0.9$ when the particle impact velocity is slow. Noting that the impact velocities in the experiments are generally below 20cm/s, we believe the use of a constant $e = 0.9$ results in an underestimate of the energy dissipation, as observed in Figure 8. To investigate the effect of a velocity-dependent $e$ on the granular flow cooling process, the new energy dissipation rate Eq. (27) is used in our improved KT models. As shown in Figure 9, there is a significant drop on the granular temperature predicted by both Model I and Model II. This shows the use of a velocity-dependent $e$ can be crucial in improving the accuracy of KT models for granular flows. Also the fact that both Model I and Model II predict similar results may indicate that in this particular free cooling case, the ratio $\lambda$ has an insignificant impact on the dissipation rate of the translational fluctuation energy, and the assumption of small rotational fluctuation energy dissipation rate used in Model I is reasonable.

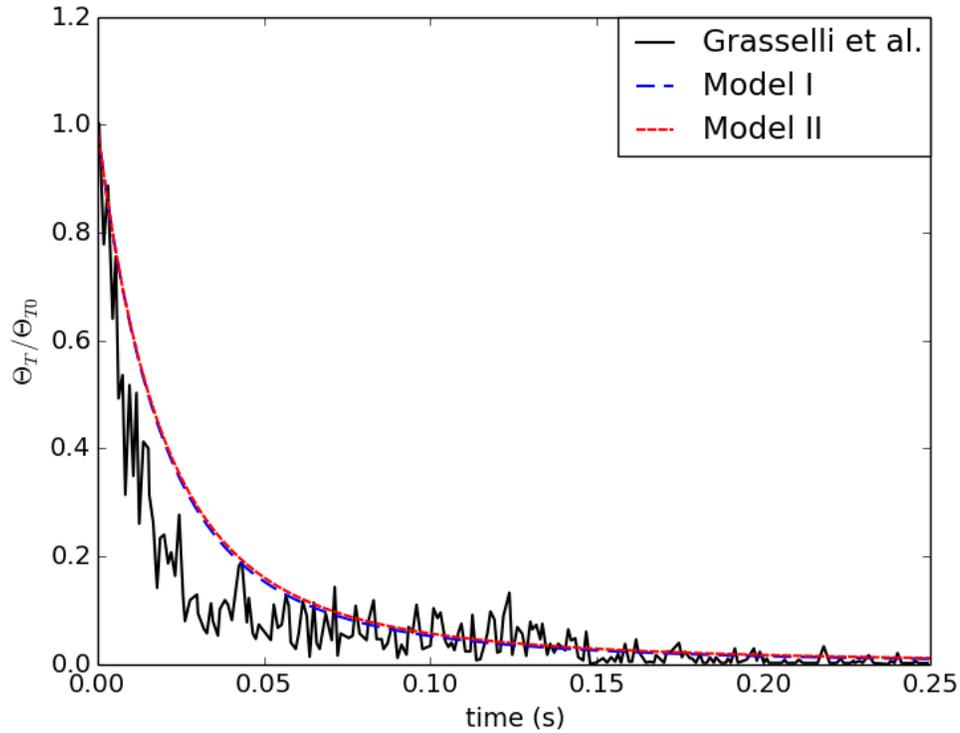

FIG. 9. Granular temperature along with the time during the free cooling process. The solid line represents the experimental data from Grasselli, Bossis and Goutallier [31]. The two dashed lines show the theoretical predictions of translational granular temperature from Model I and Model II with the velocity-dependent $e$, respectively.

However, the modified KT models still slightly overestimate the translational granular temperature, especially at the initial stage. As explained in the previous section, such discrepancy could be caused by the existence of small gap between particles and the glass walls, and the recent finding that the translations and rotations are correlated when particles are rough could also contribute to the error since the Maxwellian distribution function used here will be no longer accurate. Nevertheless, the results in Figure 9 clearly show that our improved KT models which incorporate the velocity-dependent restitution coefficient $e$ and the frictional effect are able to lower the predicted granular temperature and produce results that agree well with the experimental measurements.

## V. CONCLUSION

Previous modifications of the KT models for granular flows are limited to small restitution coefficients and result in a reduced energy dissipation rate that is not physically possible at large friction coefficient. These models could significantly underestimate the decay rate of the granular temperature in the system during the free cooling process. To improve the accuracy of current theory and facilitate the modeling of a wide range of granular flow systems, we have developed improved KT models that are able to incorporate the particle surface friction and the increased inelasticity at small impact velocities. By fitting the experimental data found in the literature, a velocity-dependent normal restitution coefficient profile has been derived and used in the present

KT models. Two different approaches that considers the particle surface friction, named as Model I and Model II, were examined and evaluated. Model I simply absorbs the effect of particle rotation into the effective restitution coefficient and only translational granular temperature is solved; Model II solves the coupled equations for both rotational and translational granular temperatures, resulting in a better accuracy. For this free cooling case, both Model I and Model II could predict similar results with good accuracy. Comparing with the experimental results, we have shown that both the velocity-dependent restitution coefficient and the particle surface friction are important in accurately predicting the granular temperature of the system; our KT models that integrate these two factors are able to improve the simulation results and produce a good agreement with the experimental measurement.

## APPENDIX

For 3D systems the unperturbed particle distribution function could be written as

$$f(c,\omega,r,t) = n\left(\frac{1}{2\pi\Theta_T}\right)^{3/2}\left(\frac{I}{2\pi m\Theta_R}\right)^{3/2}\exp\left(-\frac{I\omega^2}{2m\Theta_R}-\frac{c^2}{2\Theta_T}\right) \tag{A1}$$

The part of the translational fluctuation energy dissipation that is related to the normal restitution coefficient in homogeneous cooling state is given by

$$\Gamma_{T1} = \frac{d^2}{2}\int_{c_{12}\cdot k<0}\frac{1}{4}mg_0\left[1-e_0^2\left(1-A\exp\left(-\frac{(c_{12}\cdot k)^2}{V_a^2}\right)\right)\right](c_{12}\cdot k)^3 \tag{A2}$$

$$n^2\left(\frac{1}{2\pi\Theta_T}\right)^3\left(\frac{I}{2\pi m\Theta_R}\right)^3\exp\left(-\frac{I(\omega_1^2+\omega_2^2)}{2m\Theta_R}-\frac{(c_1^2+c_2^2)}{2\Theta_T}\right)dkdc_1dc_2d\omega_1d\omega_2$$

Since the energy dissipated due to the inelastic interactions in the normal direction is independent of the angular velocities, the integration of the above expression over $d\omega_1 d\omega_2$ would yield to

$$\Gamma_{T1} = \frac{d^2}{2} \int_{c_{12}\cdot k<0} \frac{1}{4} mg_0 \left[1 - e_0^2 \left(1 - A\exp\left(-\frac{(c_{12}\cdot k)^2}{V_a^2}\right)\right)\right] (c_{12}\cdot k)^3 \tag{A3}$$

$$n^2 \left(\frac{1}{2\pi\Theta_T}\right)^3 \exp\left(-\frac{(c_1^2 + c_2^2)}{2\Theta_T}\right) dk\, dc_1\, dc_2$$

To obtain the integrations in Eq. (A3), we need to transform the integral variables $dc_1 dc_2$ to $dc_{12} dc'_{12}$ where $c_1 = \frac{c'_{12} + c_{12}}{2}$ and $c_2 = \frac{c'_{12} - c_{12}}{2}$. The Jacobian of this transformation is $1/8$.

$$\Gamma_{T1} = \frac{d^2}{64} mg_0 \int_{-\infty}^{\infty} \int_{-\infty}^{\infty} \int_{c_{12}\cdot k<0} \left[1 - e_0^2 \left(1 - A\exp\left(-\frac{(c_{12}\cdot k)^2}{V_a^2}\right)\right)\right] (c_{12}\cdot k)^3 \tag{A4}$$

$$n^2 \left(\frac{1}{2\pi\Theta_T}\right)^3 \exp\left(-\frac{(c_{12}^2 + c_{12}'^2)}{4\Theta_T}\right) dk\, dc_{12}\, dc'_{12}$$

We obtain

$$\Gamma_{T1} = -12 \left[1 - e_0^2 + Ae_0^2 V_a^4 \left(\frac{2}{(4\Theta_T + V_a^2)^2} - \frac{A}{(8\Theta_T + V_a^2)^2}\right)\right] \frac{\rho g_0 \phi^2}{d\sqrt{\pi}} \Theta_T^{3/2} \tag{A5}$$

By treating $\theta_c$ as constants and $\beta_f$ independent of the impact velocity, expression for the decay rate of translational fluctuation energy would be

$$\Gamma_T = -12 \left[1 - e_0^2 + Ae_0^2 V_a^4 \left(\frac{2}{(4\Theta_T + V_a^2)^2} - \frac{A}{(8\Theta_T + V_a^2)^2}\right)\right] \frac{\rho g_0 \phi^2}{d\sqrt{\pi}} \Theta_T^{3/2} \tag{A6}$$

$$- 48 \left[\frac{\eta_0}{2}\left(\frac{\arctan\sigma}{\sigma} + \frac{1}{\sigma^2+1}\right) - \frac{1}{\sigma^2+1}(\lambda+1)\eta_0^2\right] \frac{\rho g_0 \phi^2}{d\sqrt{\pi}} \Theta_T^{3/2}$$

Similarly, for 2D systems we have

$$\Gamma_T = -4\left[1 - e_0^2 + Ae_0^2 V_a^4 \left(\frac{2}{\left(4\Theta_T + V_a^2\right)^2} - \frac{A}{\left(8\Theta_T + V_a^2\right)^2}\right)\right] \frac{\rho g_0 \phi^2}{d\sqrt{\pi}} \Theta_T^{3/2} \quad \text{(A7)}$$

$$-16\left[\frac{\eta_0}{2}\frac{1+\sigma^2}{\left(1+\sigma^2\right)^{3/2}} - \frac{\eta_0^2}{2}\frac{(1+\lambda)}{\left(1+\sigma^2\right)^{3/2}} - \eta_0^2 \tan^2\theta_c \left(1 - \frac{1+\frac{3}{2}\sigma^2}{\left(1+\sigma^2\right)^{3/2}}\right)\right] \frac{\rho g_0 \phi^2}{d\sqrt{\pi}} \Theta_T^{3/2}$$